\documentclass[12pt,fleqn]{article}
\begin{document}

\title{Rigidly Rotating Strings in Stationary Axisymmetric Spacetimes}
\author{\\
V. Frolov${}^{1,2,3}$\, ,
S. Hendy${}^{1}$ and J.P. De Villiers${}^{1}$}
\maketitle
\noindent
$^{1}${\em Theoretical Physics Institute, Department of Physics,
\ University of Alberta, Edmonton, Canada T6G 2J1}
\\ $^{2}${\em CIAR Cosmology Program}
\\ $^{3}${\em P.N.Lebedev Physics Institute,  Leninskii Prospect 53, Moscow
117924, Russia}

\maketitle
\noindent

\newpage
\section*{Abstract}

In this paper we study a motion of a rigidly rotating
Nambu-Goto test string in a stationary axisymmetric background spacetime.
As special examples we consider rigid rotation of strings in a flat spacetime,
where explicit analytic solutions can be obtained, and in the Kerr spacetime
where we find an interesting new family of test string solutions. We present
a detailed classification of these solutions in the Kerr background.

\newpage
\section{Introduction}

Cosmic strings are cosmologically interesting objects \cite{Vilenkin} and the
motion of strings in a curved background is a subject which has recently been
intensively discussed \cite{Sanchez, Veneziano}.   If one neglects the
gravitational effects of a string and assumes that its thickness is zero, then
the
string configuration is a timelike minimal surface  that is an extremum of the
Nambu-Goto action. One of the interesting physical applications of the general
theory is study of the interaction of cosmic strings with a black hole.  If a
(infinitely) long cosmic string passes nearby a black hole it can be captured
\cite{Moss}. Final stationary configurations of a trapped string were analyzed
in
\cite{Us96b} and their complete analytic description was obtained.

In the general case a stationary string in a stationary spacetime is defined as
a timelike minimal surface that is tangent to the  Killing vector generating
time translations. In the Kerr-Newman metric the equations describing a
stationary string allow separation of variables \cite{Frolov89, Carter89,
Carter91}  and can be solved exactly \cite{Frolov89}.  In this paper we
generalize these results to a wider class of string configurations. Namely we
study    rigidly rotating strings in a stationary axisymmetric background
spacetime. A rigidly rotating string is a string which at different moments of
time has the same form so that its configuration at later moment of time can be
obtained by the rigid rotation of the initial  configuration around the axis of
symmetry. Denote by   $\xi_{(t)}$ and $\xi_{(\phi)}$ Killing vectors that are
generators of time translation  and rotation. The timelike minimal worldsheets
which represent a stationary rigidly rotating string are characterized by the
property that the special linear combination $\xi_{(t)}+\Omega \xi_{(\phi)}$ is
tangent to the worldsheet.  Our aim is to study such configurations in a
stationary axisymmetric spacetime.

The paper is organized as follows. General equations for a stationary rigidly
rotating string in a stationary spacetime are obtained and analyzed in
Section~2.  As the simplest application we obtain explicit analytical solutions
describing rotating strings in a flat spacetime (Section~3). One of the
interesting results is the possibility of the rigid rotation of the string with
(formally) superluminal velocity, i.e. when $r\Omega>1$ ($r$ is the distance
from the axis of rotation).  A simple explanation of this phenomenon is  given
in Section~3. Section~4 devoted to  rigidly  rotating strings in the Kerr
spacetime. To conclude Section~4 we present a classification of this new family
of test string solutions in the Kerr spacetime.

\section{General Equations}
\label{general}
Consider a stationary axisymmetric spacetime. Such a spacetime possesses at
least two commuting Killing vectors: $\xi_{(t)}$ and $\xi_{(\phi)}$.
If a spacetime is asymptotically flat the vector $\xi_{(t)}$ is singled out by
the requirement that it is  timelike at  infinity.   The vector  $\xi_{(\phi)}$
 is spacelike at infinity and it is
singled out by the property that its integral curves are closed lines.
The metric for a stationary axisymmetric spacetime can be written in the form
\begin{equation}
\label{0.1}
ds^2=-V \left[dt-w d\phi\right]^2 + {1 \over V} \left[\rho^2 d \phi^2 + e^{2
\gamma}
(d \rho^2 +
dz^2 )\right],
\end{equation}
where $V$, $w$ and $\gamma$ are functions of the coordinates $\rho$
and $z$ only. This is the so called the Papapetrou form of the metric for
stationary axisymmetric spacetimes (see Ref. \cite{Wald84} for example). In
these coordinates $\xi_{(t)}^{\mu}=\delta_{t}^{\mu}$ and
$\xi_{(\phi)}^{\mu}=\delta_{\phi}^{\mu}$.

Denote by $\Sigma$ a two-dimensional timelike minimal surface representing the
motion of a string in this spacetime and denote by $S_t$ the spatial slice
$t=\mbox{const}$. The intersection  of $\Sigma$ with the surface $S_t$ is a
one-dimensional line $\gamma_t$ representing the string configuration at the
time $t$. We define a rigid cosmic string as one  whose shape and extent (but
not necessarily position) are independent of the coordinate time $t$. If $x^i$
are spatial coordinates (for metric (\ref{0.1}) $(\rho, z, \phi)$) then
$\gamma_t$ is given by the equations $x^i=x^i(\sigma, t)$, where $\sigma$ is a
parameter along the string. Since $\xi_{(\phi)}$ is tangent to  $S_t$ it
is a
generator of symmetry transformations (spatial rotations) acting on $S_t$. It
is evident that this transformation preserves the form and the shape of the
string $\gamma_t$. Our assumption that the string at the moment $t$ is obtained
by a rigid rotation from the string $\gamma_{t_0}$  can be written as
\begin{equation}
\rho(\sigma, t)=\rho(\sigma, t_0)\, ,\hspace{.3cm}z(\sigma, t)=z(\sigma, t_0)\,
,\hspace{.3cm}\phi(\sigma, t)=\phi(\sigma, t_0)+f(t,t_0) \, .
\end{equation}
Moreover we assume uniform rotation, so that $f(t,t_0)=\Omega
(t-t_0)$, where  $\Omega$ is a constant angular velocity. It is evident that
the following  combination $\chi^{\mu} = \xi^{\mu}_{(t)} +  \Omega
\xi^{\mu}_{(\phi)}$ of  the Killing vectors  $\xi^{\mu}_{(t)}$ and
$\xi^{\mu}_{(\phi)}$  is tangent to the worldsheet $\Sigma$ of a uniformly
rotating string.

In a region where $\chi^{\mu}$ is timelike one can define a set of Killing
observers whose four-velocities are $u^{\mu}=\chi^{\mu}/\left| \chi^2
\right|^{1/2}$. This set of observers form a rigidly rotating reference frame
that is the frame moving  with a constant angular velocity $\Omega$.  One
could  choose to define a rigidly rotating string as a string which was fixed
in form and  position in the frame of some Killing observer with angular
velocity $\Omega$. It can be shown that if  all the string is located in the
region where $\chi^{\mu}$ is timelike this definition is equivalent to that
given above. But, as we shall demonstrate later, situations are possible when
a rigidly rotating string lies in a region where the Killing vector
$\chi^{\mu}$ is spacelike yet its world sheet surface $\Sigma$ remains
timelike.
With this possibility in mind we will use the former definition of the rigid
string rotation.

We begin by performing the following coordinate transformation:
\begin{equation}
\label{0.2}
\varphi= \phi - \Omega t,
\end{equation}
where $\Omega$ is a constant. Metric (\ref{0.1}) now  takes the form
\begin{equation}
\label{0.3}
\nonumber ds^2   =  -V [(1-\Omega w)dt-w d\varphi]^2 +
{1 \over V} \left[\rho^2 (d\varphi+\Omega dt)^2 +
e^{2 \gamma} (d \rho^2 + dz^2 )\right]
\end{equation}
and the Killing vector $\chi$ has components $\chi^{\mu}=(1,0,0,0)$. The
Killing trajectories of $\chi$ (that might be timelike or spacelike) are:
$\rho, z, \varphi=$const .

A configuration of a test cosmic string in a given gravitational background is
represented by  a timelike two-surface $\Sigma$ (a world sheet) that satisfies
the  Nambu-Goto equations of motion. A world sheet $\Sigma$ can be described in
the parametric form $x^{\mu}=x^{\mu}(\zeta^{A})$, where $x^{\mu}$ are spacetime
coordinates, and $\zeta^{A} \;  (A=0,1)$ are the coordinates on the world
sheet. For these coordinates we shall also use the standard notation
$(\zeta^{0},\zeta^{1}) =  (\tau, \sigma)$.
The Nambu-Goto action is given by
\begin{equation}\label{0.5}
{\cal S} \left[ x^{\mu} \right]= - \mu \int{d^2 \zeta \sqrt{-G}},
\end{equation}
where $G$ is the determinant of the induced metric on the world sheet
$G_{AB} = g_{\mu \nu} x^{\mu}_{,A} x^{\nu}_{,B}$ and $\mu$ is the string
tension.

For a stationary world sheet configuration one can choose parameters $(\tau,
\sigma)$
in such a way that
\begin{equation}
\label{0.4}
x^{\mu} (\zeta^{A}) = (\tau + f(\sigma), \rho(\sigma), z(\sigma),
\varphi(\sigma)) \, ,
\end{equation}
where $f$ is some function of $\sigma$.
The determinant of the induced metric $G_{AB}$ on the world sheet is
\begin{equation}
\label{0.7}
G = \frac{e^{2 \gamma} \chi^2}{V} \left({\rho^\prime}^2+{z^\prime}^2 \right) -
\rho^2 \varphi^{\prime \; 2}.
\end{equation}
where
\begin{equation}
\chi^2 =  -V + 2 \Omega w V
+ \Omega^2 (\rho^2/V - w^2 V) \, ,
\end{equation}
Note that neither the function $f$ nor its derivative $f^{\prime}$ appear in
the action so they may be specified freely. It is convenient to choose $f$ so
that the induced metric is diagonal, i.e.
$G_{\tau \sigma}=g_{\mu \nu} x^{\mu}_{, \tau} x^{\nu}_{, \sigma}=0$. In
this case we find $f$ must be chosen to satisfy the condition
\begin{equation}
f^\prime = - \frac{V w +\Omega (\rho^2/V - w^2 V)}{\chi^2}
\;\varphi^\prime.
\end{equation}
A stationary string configuration (\ref{0.4}) provides an extremum for the
reduced
Nambu-Goto action
\begin{equation}
\label{e10}
E=\mu\int d\sigma \sqrt{\frac{e^{2 \gamma}(- \chi^2)}{V}
\left({\rho^\prime}^2+{z^\prime}^2 \right)
+\rho^2 \varphi^{\prime \; 2}}.
\end{equation}
Hence a stationary string configuration $x^i=(\rho(\sigma), z(\sigma),
\varphi(\sigma))$ is a geodesic line in a three-dimensional space with the
metric
\begin{equation}
\label{e11}
dh^2=\frac{e^{2 \gamma}(- \chi^2)}{V} \left(d{\rho}^2+{dz}^2 \right)
+\rho^2 d\varphi^{ 2} \,.
\end{equation}

The Nambu-Goto equations for a stationary rigidly rotating string are
\begin{equation}
\label{0.9a}
\partial_{\sigma} \left( \frac{\chi^2 e^{2\gamma}}{\sqrt{-G} V} \rho^\prime
\right) = {1 \over \sqrt{-G}}\left[ {1 \over 2} \frac{\partial}{\partial \rho}
\left(\frac{\chi^2 e^{2\gamma}}{V} \right)
\left({\rho^\prime}^2+z^{\prime \; 2}\right) - \rho \varphi^{\prime\;2}
\right],
\end{equation}
\begin{equation}
\label{0.9}
\partial_{\sigma} \left( \frac{ \rho^2 \varphi^\prime }{\sqrt{-G}} \right) = 0,
\end{equation}
\begin{equation}
\label{0.10}
\partial_{\sigma} \left( \frac{\chi^2 e^{2\gamma}}{\sqrt{-G} V}  z^\prime
\right)
=  {1 \over 2 \sqrt{-G}} \frac{\partial}{\partial z} \left(\frac{\chi^2
e^{2\gamma}}{V} \right)
\left({\rho^\prime}^2+z^{\prime \; 2}\right),
\end{equation}
where $G$ is given by (\ref{0.7}). Equation (\ref{0.9}) can be integrated
immediately to give
\begin{equation}
\label{0.10a}
\varphi^{\prime\;2} = \frac{-\chi^2}{V} \left(\frac{L^2 e^{2\gamma}}
{\rho^2 (\rho^2 - L^2)} \right) ({\rho^\prime}^2+z^{\prime \; 2}).
\end{equation}
Here $L$ is a constant of the integration. The constant $L$ is associated with
the $\varphi$-independence  of the Lagrangian and is related to the angular
momentum of the string. In what follows we choose $L$ to be non-negative.

These equations are invariant under reparameterization $\sigma\rightarrow
\tilde{\sigma}=\tilde{\sigma}(\sigma)$. In the region where $\rho'\ne 0$ one
can use this ambiguity to put
$\sigma=\rho$. For this choice equations (\ref{0.9a})-(\ref{0.10}) reduce to
\begin{equation}
\label{0.11}
\left(\frac{d\varphi}{d \rho}\right)^2 =
\frac{- \chi^2 L^2 e^{2\gamma}}{\rho^2 (\rho^2 - L^2) V}
\left[1+\left(\frac{dz}{d\rho}\right)^2\right],
\end{equation}
\begin{equation}
\label{0.12}
\frac{d}{d\rho} \left( \frac{\chi^2 e^{2\gamma}}{\sqrt{-\tilde{G}} V}
\frac{dz}{d\rho} \right)
 =  {1 \over 2 \sqrt{-\tilde{G}}} \frac{\partial}{\partial z}
\left(\frac{\chi^2
e^{2\gamma}}{V} \right) \left[1+\left(\frac{dz}{d\rho}\right)^2\right],
\end{equation}
where
\begin{equation}
-\tilde{G} = \frac{-G}{\rho^{\prime \; 2}} =  \frac{\rho^2 \chi^2
e^{2\gamma}}{V (L^2-\rho^2)} \left[1+\left(\frac{dz}{d\rho}\right)^2\right].
\end{equation}

The solutions represent a timelike two-surface provided the determinant
$G$ is negative definite. Thus we see that the rigidly rotating strings
are confined to regions (for $V>0$) where
\begin{equation}\label{aa}
I  \equiv {L^2 - \rho^2 \over \chi^2} > 0.
\end{equation}
When $L^2=\rho^2$ the world sheet has a turning point in $\rho$ as a function
of $\varphi$.

In general, in order to ensure rigid rotation of a string, an external
force must act on it. For example, one could assume that a string has
heavy monopoles at the end and that a magnetic field is applied to force them
to move along a circle. In this case a solution of equations
(\ref{0.11})-(\ref{0.12}) describes the motion of the string interior. In order
to escape a discussion of the details of the motion of the end points we
shall use the maximal extensions of the string solutions, continuing them until
they meet the surface where $\chi^2=0$. Since the invariant $I$ changes its
sign
at this surface, the minimal surface describing the rigidly rotating string
ceases to be timelike here. The end points of such a maximally extended string
move with the velocity of light along this surface. In this paper, for brevity,
we call such solutions ``open" strings. In what follows we restrict ourselves
to the study of such ``open" strings and do not discuss how these solutions can
arise when real forces are acting on the string or when a part of the string is
involved in a rigid rotation.

Our assumption of rigidity implies that the coordinates $\rho$ and $z$ of the
end points of the string remain fixed. Under these conditions the end points of
an ``open" string are located on the surfaces where $\chi^2=0$. In a flat
spacetime this timelike surface is a cylinder located at the radial distance
$\Omega^{-1}$ from the axis of symmetry. In the general case we shall refer  to
the  surfaces where $\chi^2=0$ as ``null cylinders". Note that if $L^2-\rho^2$
vanishes at the same point as $\chi^2$ it is possible for the world sheet to
pass through the surface $\chi^2=0$ and remain regular and timelike.

In order to find a rigidly rotating string configuration one needs to fix
functions $V(\rho,z)$, $\gamma(\rho,z)$, and $w(\rho,z)$ that specify
geometry. It is quite interesting (as was remarked by de Vega and Egusquiza
\cite{deVega96}) that if the metric (\ref{0.1}) allows a discrete symmetry
$z\rightarrow -z$, then equations (\ref{0.11}) and (\ref{0.12}) always have
a special solution, namely a string configuration described by the relations
$z=0$ and $\varphi=\mbox{const}$. De Vega and Egusquiza called these straight
rigidly rotating strings  in axially-symmetric stationary spacetimes
``planetoid" solutions.

\section{Rotating Strings in Flat Spacetime}
\label{flat}

Our main goal is a study of rigidly rotating strings in a spacetime of a
rotating black hole. But before considering this problem we make a few
remarks concerning rigidly rotating strings in a flat spacetime.  We recover
the Minkowski metric
\begin{equation}
\label{3.1}
ds^2 = - dt^2 + d\rho^2 + dz^2+\rho^2 d\phi
\end{equation}
from (\ref{0.1}) by setting the
metric functions $V=1$ and $w=\gamma=0$. We also have
 $\chi^2=\Omega^2 \rho^2 - 1$. Since the metric is independent of $z$ one
can integrate (\ref{0.12}) once to reduce the equations of motion to the form
\begin{eqnarray}\label{A.11}
{d{\varphi} \over d\rho} & = & \pm {L\,\left( 1 -
{\Omega}^{2}\,{\rho}^{2}\right) \over \rho\,\sqrt{\left( 1 - {p}^{2} +
{L}^{2}\,{\Omega}^{2}\right)\,{\rho}^{2} - {\Omega}^{2}\,{\rho}^{4} - {L}^{2}
}},\\
\label{A.12}
{d{z} \over d\rho} & = & \pm {p \,{\rho} \over \sqrt{\left( 1 - {p}^{2} +
{L}^{2}\,{\Omega}^{2}\right){\rho}^{2} - {\Omega}^{2}\,{\rho}^{4} - {L}^{2}
}},
\end{eqnarray}
where $p$ is a constant of integration. These equations can be solved
analytically.

In order for (\ref{A.11}) and  (\ref{A.12})  to be real-valued, $\rho$ is
constrained to lie in the interval $0 < {\rho}_- \leq \rho < {\rho}_+$
where the upper and lower bounds are given by,
\begin{equation}\label{A.00}
\rho_{\pm} = \Omega^{-1} \sqrt{(B\pm C)/ 2} \, ,
\end{equation}
where,
\begin{equation}\label{A.17b}
B = 1 - {p}^{2} + {L}^{2}\,{\Omega}^{2}, \mbox{ and } \,
{C} = \sqrt{{B}^{2} - 4\,{\Omega}^{2}\,{L}^{2}} \, .
\end{equation}
The equations for $ z(\rho)$ and $\varphi(\rho)$ can be integrated readily
(the substitution $ u = {\rho}^{2}$ reduces these to standard integrals) with
solutions,
\begin{eqnarray}\label{A.17a}
z_{\pm}(\rho)&=&\mp {p \over 2 \Omega}\,\left(
\arcsin{{B-2\,{\Omega}^{2}\,{\rho}^{2}  \over C}} -{\pi \over 2}\right),\\
\nonumber \varphi_{\pm}(\rho) &=& \pm \,{1 \over 2}\left\{ \arcsin{
{B\,{\rho}^{2}-2\,{L}^{2}  \over C\,{\rho}^{2} } } \right.\\
\label{A.0} &\hspace{0.5cm}& + \left. \Omega\,L\, \arcsin{
{B-2\,{L}^{2} \,{\rho}^{2} \over C } } +{\pi \over 2}\,\left( 1-
\Omega \, L \right) \right\}
\end{eqnarray}
where for convenience we have chosen the initial conditions
$z_{\pm}(\rho_-)=\varphi_{\pm}(\rho_-)=0$.

It is instructive to examine the special case where $p = 0$ further where the
solutions are confined to the $z =$ const plane.
The solution (\ref{A.0}) can be rewritten in the form
\begin{equation}\label{A.16}
\varphi_{\pm}(\rho) = \pm \,\left( \,\arctan{k\,\xi} -
k^{-1} \arctan{\xi} \right) \, ,
\end{equation}
where $k= ( \Omega L)^{-1}$ and
$\xi= \Omega \sqrt{(\rho^{2} - L^{2})/ (1 - \Omega^2 \rho^{2})} $.

In order for solutions to exist, the invariant $I \equiv {(L^2-\rho^2)/
\chi^2}$  must be non-negative. Thus there are a number of cases to resolve. We
know that  the string can end only on the null cylinder where $\chi^2$
vanishes, i.e. $\rho = 1/ \Omega$. We also see that the string may have a
turning point at  $\rho = L$. When $L=0$ we recover the rigidly rotating
straight strings of De Vega and Egusquiza \cite{deVega96}. When $L > 0$ there
are two cases;
\begin{enumerate}
\item $L < 1/\Omega$: the string lies in the region $L < \rho <1/\Omega$, has
end-points at $\rho=1/\Omega$ and a turning point at $\rho=L$ (see Figure~1),
\item $L > 1/\Omega$: the string lies in the region $L > \rho >1/\Omega$. It
has
end-points at $\rho=1/\Omega$ and a turning point at $\rho=L$ (see Figure~2).
\end{enumerate}
(The case $L=1/\Omega$ is excluded since $I<0$ and hence  no solution exists.)

In the latter case the Killing vector $\chi$ is spacelike. Nonetheless the
world-sheet is timelike; in fact the tangent vector $x^{\mu}_{,\sigma}$ is
timelike in this region. However the solution lies in the region $L > \rho
>1/\Omega$  and appears, by comparing $t=\mbox{const}$ slices in non-rotating
coordinates, to move at ``superluminal velocities" (except at the end points
which move at  the speed of light). This is in apparent contradiction with the
observation that the world sheet is timelike.

The puzzle is clarified if we note that the apparent velocity of the string in
the surface $t=\mbox{const}$ is not the physical velocity of the string. Recall
that the Nambu-Goto action is invariant under world-sheet reparameterizations.
This reparameterization can be used to generate a ``motion" of the string along
itself, which evidently is physically irrelevant.  In other words only
velocity  normal to the string world-sheet has physical  meaning (see e.g.
\cite{Vilenkin}).

Hence, on the $t=\mbox{const}$ hypersurfaces, we must to consider the component
of the apparent string velocity {\em normal} to the string configuration. The
normal component of the velocity is the physical component. The
apparent three velocity,  $v^i\;(i=1,2,3)$, of the string in $(\rho, z, \phi)$
coordinates as measured by a static observer at infinity is
\begin{equation}
v^i=\Omega (0,0,1)
\end{equation}
and has magnitude $v= \rho \Omega$.

Its projection on the normal to the string $u^i_{\bot}$ in the
$(t=\mbox{const})$ plane is
\begin{equation}
u^i_{\bot}=\frac{\Omega}{1+\rho^2 \varphi^{\prime \; 2}}
(-\rho^2 \varphi^{\prime},0,1),
\end{equation}
and the magnitude of this normal velocity is
\begin{equation}
u^2_{\bot}=\frac{\rho^2 \Omega^2}{1+\rho^2 \varphi^{\prime \; 2}}
= \frac{\Omega^2 (L^2-\rho^2)}{\Omega^2 L^2 - 1}.
\end{equation}
Thus we see that if $1/\Omega^2 > \rho^2 > L^2$ (case 1) then $u^2_{\bot} < 1$
as expected. Furthermore if $L^2 > \rho^2 > 1/\Omega^2$ (the apparently
``superluminal" case 2) we see that $u^2_{\bot} < 1$ also. The physical
velocity of the string is subluminal in all cases where the solution exists.

This phenomenon is evidently of a quite general nature. In order to separate
these two different types of rigid rotation of strings we will call the motion
``superluminal" if they are tangent to $\chi$ with $\chi^2>0$ and ``subluminal"
if the world-sheet is tangent to $\chi$ with $\chi^2<0$.

\section{Rigidly Rotating Strings in the Kerr Spacetime}
\label{Kerr}

\subsection{Equations of Motion}

In Boyer-Lindquist coordinates the Kerr metric is given by:
\begin{eqnarray}
\label{1.1}
\nonumber ds^2 =& -&\frac{\Delta}{\Sigma}
\left[dt-a \sin^2\theta d\phi \right]^2 +\frac{\Sigma}{\Delta}dr^2
+\Sigma d\theta^2 \\
& + & \frac{\sin^2\theta}{\Sigma} \left[(r^2+a^2)d\phi -a dt \right]^2,
\end{eqnarray}
where  $\Delta=r^2-2Mr+a^2$ and $\Sigma=r^2+a^2\cos^2\theta$. This
spacetime is stationary and axisymmetric with  two  Killing vectors:
$\xi^{\mu}_{(t)} =\delta^{\mu}_{t}$  and $\xi^{\mu}_{(\phi)} =
\delta^{\mu}_{\phi}$.

The relationship between the Boyer-Linquist coordinate functions and the
Papapetrou coordinate functions is straightforward; the time coordinate $t$
and the angular coordinate $\phi$ that appear in both the metric (\ref{0.1})
and
the metric (\ref{1.1}) are simply identified and
\begin{equation}
\label{1.1a}
\rho^2= \Delta \sin^2\theta \, ,\;\;\;\;\;\; z = (r-M) \cos\theta.
\end{equation}
In terms of the Boyer-Lindquist coordinates the Papapetrou metric functions
for the Kerr metric are
\begin{eqnarray}
\label{1.1b}
V & = & \frac{\Delta-a^2 \sin^2\theta}{\Sigma},\\
w & = &  {2Mra\sin^2\theta \over \Delta-a^2 \sin^2\theta}, \\
e^{2 \gamma} & = & \frac{\Delta-a^2 \sin^2\theta}{\Delta \cos^2\theta +
(r-M)^2 \sin^2\theta}.
\end{eqnarray}

We can now write down the string equations of motion for the Kerr
spacetime in the Boyer-Linquist coordinates. The string configuration is
determined by two functions $\varphi(r)$ and $\theta (r)$ which satisfy
\begin{equation}
\label{1.9}
\left( \frac{d\varphi}{dr}\right)^2
=  {\tilde{G} L^2\over \Delta^2 \sin^4\theta}
\end{equation}
\begin{equation}
\label{1.10}
 {\Sigma \chi^2 \over \sqrt{\tilde{G}}} \frac{d}{dr}
\left( {\Sigma \chi^2 \over \sqrt{\tilde{G}}} \frac{d\theta}{dr} \right)
 =  { \cos\theta \over \Delta \sin^3\theta} Z\, ,
\end{equation}
where
\begin{equation}
Z= L^2+ b^2\left(1+{\Delta (1- a^2 \Omega^2
\sin^2\theta) \over \Sigma \chi^2} \right)
\end{equation}
\begin{equation}
b^2=L^2-\Delta \sin^2\theta \, ,\hspace{0.5cm}
\tilde{G}= {\Sigma \chi^2 \sin^2\theta \over{b^2} }
\left[1 + \Delta \left(\frac{d \theta}{dr}\right)^2\right].
\end{equation}
\begin{equation}
\chi^2=\frac{\sin^2\theta \left[a^2+4\Omega M r a +
\Omega^2((r^2+a^2)^2-\Delta a^2 \sin^2\theta) \right] - \Delta}{\Sigma}
\end{equation}

We were not able to solve this system analytically. Moreover one
can show (see Appendix) that the only case when the variables in the problem
under consideration can be separated by the Hamilton-Jacobi method (at least in
these coordinates) is when $\Omega=0$, so that the problem reduces to the one
considered in \cite{Frolov89}.  In what follows we consider the case when a
rotating string is located in the equatorial plane, which allows more detailed
analysis.

\subsection{Rigidly Rotating Strings in the Equatorial Plane}
\label{sub4.2}

For the motion of the string  in the
equatorial plane $\theta=\pi/2$ equation
(\ref{1.10}) is satisfied  identically (both the left and right hand sides
vanish) and equation (\ref{1.9}) takes the form
\begin{equation}
\label{2.1}
\left(\frac{d\varphi}{dr}\right)^2 =  {\Sigma L^2 \chi^2
\over \Delta^2 (L^2-\Delta) }\, .
\end{equation}
Here
\begin{equation}
\label{2.1a}
\chi^2 ={F\over r}\,  ,\hspace{.5cm}F=\Omega^2
r^3+(a^2\Omega^2-1)r+2M(a\Omega-1)^2\, .
\end{equation}
Solutions exist only if the right-hand
side of (\ref{2.1}) is non-negative and hence (for $L\ne 0$)
\begin{equation}
\label{2.2}
I\equiv {L^2 - \Delta \over \chi^2} \geq 0.
\end{equation}
The positivity of the invariant $I$ also guarantees that the world-sheet of
the string is  a regular timelike surface (cf. (\ref{aa})).

To simplify the analysis of the structure of the null cylinder surfaces instead
of $r$, $a$, $L$,$\Omega$ and $F$ we introduce the dimensionless variables
\begin{equation}\label{ee}
\rho={r\over M}\, ,\hspace{0.5cm}\alpha={a\over M}\,
,\hspace{0.5cm}\lambda={L\over M}\,
,\hspace{0.5cm}\omega=M\Omega\,,\hspace{0.5cm}$F=Mf$\,.
\end{equation}
In these variables
\begin{equation}\label{ff}
\chi^2 ={f\over \rho}\,  ,\hspace{.5cm}f=\omega^2
\rho^3+(\alpha^2\omega^2-1)\rho+2(\alpha\omega-1)^2\, . \end{equation}

The surface where $L^2 - \Delta$ vanishes
corresponds, in general, to turning points of the string in $r$. Now
$L^2-\Delta$ has only one zero outside the
horizon, namely  $r_0 = M\rho_0$ with
\begin{equation}\label{}
 \rho_0=1+ \sqrt{1+\lambda^2-\alpha^2}\, .
\end{equation}
For $r > r_0$, $L^2-\Delta < 0$,  and for $r < r_0$, $L^2-\Delta >0$.

Firstly we note that when $L=0$ then we obtain the planetoid solution of
De Vega and Egusquiza \cite{deVega96}. In this case the solution is a rigidly
rotating straight string with end points on the null cylinders $r=r_1$ and
$r=r_2$ where $r_2 > r_1$ are the zeroes of $\chi^2$. Note that for a given
$\Omega$ if these zeroes do not exist (so that $\chi$ is spacelike everywhere)
then there are no such rigidly rotating {\em straight} strings.

In the general case the endpoints of an ``open" string must be located on the
null
cylinders  where $\chi^2=0$. Note one can think of the equation
$f(\rho,\omega, \alpha)=0$ as a quadratic in $\omega$ for fixed $\rho$ and
$\alpha$.
The zeroes of (\ref{ff}), that is the solutions of the equation
$f(\rho,\omega,\alpha)=0$, are
\begin{equation}
\label{2.4}
\omega_\pm = {2  \alpha  \pm \rho \sqrt{\rho^2-2\rho+\alpha^2} \over \rho^3
+\alpha^2 \rho +2\alpha^2}.
\end{equation}
They bound the interval $\omega_-(\rho)<\omega<\omega_+(\rho)$ where $\chi$ is
timelike at a given radius $r=M\rho$ . We note that inside the horizon where
$\rho^2-2\rho+\alpha^2 < 0$, it is not possible for $\chi$ to be timelike or
null except within or on the inner Cauchy horizon. Since we are interested in
the motion of the strings in the black hole exterior from now on we restrict
ourselves by solutions in the region $\rho>\rho_+$ where

\begin{equation}
\label{rhoplus}
\rho_+=1+\sqrt{1-\alpha^2}.
\end{equation}

Equation (\ref{2.4}) shows that $\omega_{\pm}|_{r_+}=\omega_{BH}\equiv
\alpha/(\rho_+^2+\alpha^2)$ ($\omega_{BH}/M$ is the angular velocity of the
black hole).  At large distances $\omega_{\pm}=\pm 1/\rho$ reproduces flat
space behavior. For fixed $\alpha$ the function $\omega_+(\rho,\alpha)$ has a
maximum, and the function $\omega_-(\rho,\alpha)$ has a minimum. The points of
the extrema can be defined as joint solutions of the equation
$f(\rho,\omega,\alpha)=0$ and the equation $\left. {\partial f/ \partial
\rho}\right|_{(\omega,\alpha)}=0$. The latter equation implies that
\begin{equation}\label{cc}
\rho=\pm \rho_m\, ,\hspace{.5cm} \rho_m={\sqrt{1-\omega^2 \alpha^2}\over
\sqrt{3}|\omega|}\, .
\end{equation}

A simultaneous solution of this relation and equation (\ref{2.4}) defines a
maximal value $\omega_{max}(\alpha)$ of $\omega_+$ and a minimal value
$\omega_{min}(\alpha)$ of $\omega_-$. We conclude that for a given value of the
rotation parameter $\alpha$
the Killing vector $\chi$ can be timelike only if
$\omega_{min}(\alpha)<\omega<\omega_{max}(\alpha)$. If there exists a region
where the Killing vector $\chi$ is timelike, this region is inside interval
$\rho_1<\rho<\rho_2$, and one has $\rho_{min}(\alpha)<\rho_1$ and
$\rho_{max}(\alpha)>\rho_2$ (where $\rho_1(\omega,\alpha) <
\rho_2(\omega,\alpha)$
are the zeroes of the polynomial $f$). Here $\rho_{min}(\alpha)$ and
$\rho_{max}(\alpha)$ are given by (\ref{cc}) with $\omega=\omega_{min}(\alpha)$
and $\omega=\omega_{max}(\alpha)$, respectively.

We can arrive at the same conclusion by slightly different reasoning
which will allow us to make further simplifications. For fixed $\alpha$ and
$\omega$  the
function $f(\rho,\omega,\alpha)$  is a cubic polynomial in $\rho$ . It tends to
$\pm \infty$ as $\rho\rightarrow\pm\infty$
and $f(0)\ge 0$ ($f(0)=0$ only if   $\omega \alpha= 1$). For
$\alpha|\omega|\ge1$ it is monotonic, and hence always positive at $\rho>0$.
For $\alpha|\omega|<1$ the function $f(\rho)$ has a  minimum  at $\rho=\rho_m$
and a maximum at $\rho=-\rho_m$, where $\rho_m$ is given by (\ref{cc}). At the
minimum point, $f$ takes the value \begin{equation}\label{hh}
f_m=f(\rho_m,\omega)=2 (\omega \alpha-1)^2\left[ 1-{1\over
3\sqrt{3}|\omega|}{\sqrt{(1+\omega \alpha)^3\over 1-\omega \alpha}}\right]\, .
\end{equation}
The minimum value $f_m$ vanishes if the following equation is satisfied
\begin{equation}
\label{nn}
\alpha(\alpha^2+27)\omega^3+(3\alpha^2-27)\omega^2+3\alpha \omega+1=0\,.
\end{equation}
Solutions of this equations $\omega(\alpha)$ are also solutions of the two
equations $f=0$ and $\partial_r f=0$, and hence they coincide with
$\omega_{min}(\alpha)$ and $\omega_{max}(\alpha)$.

The numerical solution of  equation (\ref{nn}) is shown in Figure~3. Line $a$
represents solution $\omega_{max}$ and line $b$ represents solution
$\omega_{min}$. These lines begin at $\alpha=0$ at their Schwarzschild values
$\pm 3^{-3/2}$ and reach values
 $1/2$ and $-1/7$ respectively for the extremely rotating black hole.  Figure~4
shows the
corresponding radii $\rho_{max}$ (curve a) and $\rho_{min}$ (curve b) as the
functions of the
rotation parameter $\alpha$.

The third branch $c$ in Figure~3, which intersects $a$ at $\alpha=1$,
corresponds to the minimum values of $\omega$ inside the Cauchy horizon.
Line $d$  is the solution of the equation $\omega\alpha=1$. For the values
of the parameters in the $(\alpha-\omega)$ plane lying in  the region outside
two shaded strips function $f
$ is positive for any $\rho>0$.  For the values of the parameters inside two
shaded strips function $f$ has
two roots, $0<\rho_1<\rho_2$, and $f$ is negative for $\rho$ lying between the
roots. The upper shaded region corresponds to the situation where the roots of
$f$ lie within the inner Cauchy horizon. The lower shaded region is the area
where the two roots
of $f$ lie outside the event horizon.

Having obtained this information on the structure of the null cylinder surfaces
we now discuss the different types of motion of a rigidly rotating string in
the equatorial plane of the Kerr spacetime. The simplest situation clearly
occurs when $\chi$ has no zeroes in the region $\rho \ge \rho_+ $ (where
$\rho_+$ is defined by equation (\ref{rhoplus})).  This happens for values of
the parameters which lie outside the shaded region restricted by lines $a$ and
$b$
in the $(\alpha-\omega)$-plane (see Figure~3). In this case there is only one
allowed type of solution $\rho_0>\rho_+$ corresponding to solutions in the
region $\rho_+ \leq \rho \leq \rho_0$.  These configurations begin and end in
the black hole and have a turning point at $\rho=\rho_0$ in its exterior (the
form of these solutions is qualitatively similar to that of the solution
shown in figure~6). Such a solution may describe a closed loop-like string,
part
of which has been swallowed by a black hole. Centrifugal forces connected with
the rotation allow the other part of the string to remain in the
black hole exterior.

For the values of the parameters lying inside the shaded region restricted by
lines $a$ and $b$ in the $(\alpha-\omega)$-plane the situation is somewhat more
complicated.  In this case $\chi^2$ has two zeroes that we denote by $\rho_1$
and $\rho_2$ ($\rho_+ < \rho_1 <\rho_2$). The Killing vector $\chi$ is timelike
in the region $\rho_1<\rho<\rho_2$. Since the invariant $I$ defined by
(\ref{2.2})
must be positive there are then a number of different possibilities depending
upon the choice of the angular momentum parameter $\lambda\ge 0$.

\begin{enumerate}
\item $\rho_0<\rho_1$. The invariant $I$ is positive either if $(a)$ \ \
$\rho_1<\rho<\rho_2$ or $(b)$\ \ $\rho<\rho_0$. In the former case $f<0$
and the motion of the string is ``subluminal", with the ends of the string
at $\rho_1$ and $\rho_2$ (Figure~5). In the latter case the motion of the
string is ``superluminal",
the string begins and ends on the black hole and has a  radial turning point
$\rho_0$ in the black hole exterior (Figure~6).
\item $\rho_1<\rho_0<\rho_2$. The invariant $I$ is positive either if
$(a)$ \ \    $\rho_0<\rho<\rho_2$ or $(b)$\ \  $\rho<\rho_1$. In the
former case $f<0$ and the motion of the string is ``subluminal", with both ends
of the string at $\rho_2$ and $\rho_0$ being radial turning points (two
examples (with $\lambda=0.5$ and $\lambda = 2.0$) are shown in Figure~7). In
the
latter case the motion of the string is ``superluminal" and string begins at
the
horizon
$\rho_+$ and ends at $\rho_1$ (Figure~8).
\item $\rho_2<\rho_0$. The invariant $I$ is positive either if  $(a)$ \ \
 $\rho_2<\rho<\rho_0$ or $(b)$\ \  $\rho<\rho_1$. In both cases $f>0$ and the
motion is  ``superluminal". In the former case  the ends of the string are at
$\rho_2$ and a radial turning point is at $\rho_0$ (an example of such a
configuration with $\lambda=4.0$ is given
in Figure~7).  In the latter case the string configurations are similar to the
one shown in Figure~8. \end{enumerate}

For ``subluminal" motion the apparent velocity is less than the velocity of
light, for ``superluminal" motion the apparent velocity is greater than the
velocity of light. In both cases the physical (orthogonal to the string)
velocity is less than the
velocity of light. We described the origin of this phenomenon in Section~3.
We recall that in the above analysis we restricted ourselves to the case of
rotating strings. In the absence of rotation stationary string
configurations (both equatorial and off-equatorial) allow a complete
description (see \cite{Frolov89}).

Besides these main categories of motion there are possible different boundary
cases when $\rho_0$ coincides either with $\rho_1$ or with $\rho_2$. These
cases require special analysis. For special values of the parameters one might
expect that a string passes through (and beyond) these points remaining regular
and timelike (see Figure~9 for example). A similar situation was analysed in
\cite{Us96b} for special configurations of strings which pass through the event
horizon.

\section*{Acknowledgements}
The work of V.F. and J.P.D. was supported by NSERC, while the work by S.H. was
supported by the International Council for Canadian Studies. The authors would
like to thank Arne Larsen for useful discussions.

\appendix
\section*{Appendix}

In this appendix we analyze separability of the Hamilton-Jacobi equations
describing rigidly rotating strings in the Kerr spacetime.
Since a rigidly rotating string in the axisymmetric stationary spacetime
provides the minimum of the energy functional (\ref{e10}) and its configuration
is a geodesic in a three-dimensional space with metric  (\ref{e11}) one can use
the following form of the Hamilton-Jacobi equation for defining such
configurations
\begin{equation}\label{A.5}
{\partial {S}\over \partial \sigma}+ {1 \over 2} {h}_{i j}  {\partial {S}\over
\partial {x}^{i}} {\partial {S}\over \partial {x}^{j}} = 0\, .
\end{equation}
Here $h^{ij}$ is the metric inverse to $h_{ij}$ given by  (\ref{e11}).

For the special case of
Minkowski spacetime, where $V=1$, $w=\gamma=0$, (and setting $m=1$), the
Hamilton-Jacobi equations are trivially separable
\begin{equation}\label{A.6}
S = - {1\over 2} \,{{m^2}\,\sigma }+  L\,\varphi  + pz+ A(\rho ) \, ,
\end{equation}
and function $A(\rho)$ obeys the equation
\begin{equation}\label{A.8a}
\left({{d{A(\rho)} \over d\rho}}\right)^{2} + p^2 +
\left( {L \over {\rho}^{2}} - 1\right)\,\left(1 - {\rho}^{2}\,{\Omega}^{2}
\right)
 = 0\, .
\end{equation}
By solving these equations and varying the action $S$ with respect to
separation constants and $m$ one gets the complete set of equations that is
equivalent to  (\ref{A.11}) and  (\ref{A.12}).

Turning now to the case of Kerr spacetime, it is easier to apply transformation
(\ref{0.2}) directly to the Kerr metric (\ref{1.1}) and repeat the steps
outlined above. It is straightforward, again, to show that the components of
the spatial metric are,
\begin{equation}\label{A.13a}
{h}_{r r} =\Delta^{-1}{h}_{\theta \theta}\, ,\hspace{.5cm}
{h}_{\theta \theta} =  \Delta\,{T}^{2} -{
\sin}^{2}{\theta}\,{R}^{2}\, ,\hspace{.5cm}{h}_{ \varphi\varphi} =  \Delta\,{
\sin}^{2}{\theta}
\end{equation}
where,
\begin{equation}\label{A.13d}
T(\theta) = 1 - a\,\Omega\,{ \sin}^{2}{\theta}
\, ,\hspace{.5cm}R(r) = a - \Omega\,({r}^{2} + {a}^{2})\, .
\end{equation}

Assuming the separation of variables in this metric and working  with an action
of the form,
\begin{equation}\label{A.14a}
S = - {1\over 2} \,{{m^2}\,\sigma }+  L\,\varphi  + A(r) + B(\theta)
\end{equation}
the Hamilton-Jacobi equation yields
\begin{eqnarray}
\nonumber  & \Delta & \left({\,{d{A(r)} \over dr}}\right)^{2}  +
\left({{d{B(\theta)} \over
d\theta}}\right)^{2} +
 {{L}^{2}\,{T}^{2} \over { \sin}^{2}{\theta}}  \\ &-&  {{L}^{2}\,{R}^{2}
\over\Delta}
\label{A.15a}  -  {m}^{2}\,\left[{\Delta\,{T}^{2} - {R}^{2}\,{
\sin}^{2}{\theta}}\right] = 0.
\end{eqnarray}

The ${L}^{2}$ terms are effectively separated. Only the ${m}^{2}$ term requires
consideration.  In order to complete the separation of variables, it is
required
that,
\begin{equation}\label{A.16b}
 {m}^{2}\,\left[ {\Delta\,{T}^{2} - {R}^{2}\,{ \sin}^{2}{\theta}}\right]
 = {K}_{1}(r)+{K}_{2}(\theta).
\end{equation}

Expanding the bracket,
\begin{eqnarray}\label{A.16c}
&& \left[ {\Delta\,{T}^{2} -
{R}^{2}\,{ \sin}^{2}{\theta}}\right] =  \Delta - {a}^{2}\,{\sin}^{2}{\theta} \,
\\ \nonumber
&  &+ \Omega\,{ \sin}^{2}{\theta}
\left\{ \Omega\,\left( \Delta\,{a}^{2}\,{ \sin}^{2}{\theta}-{\left( {r}^{2} +
{a}^{2}\right)}^{2} \right) + 2\,a\,\left( 2\,M\,r-{Q}^{2} \right)\right\}.
\end{eqnarray}

The first two terms are separated. Simple analysis shows that the last one
cannot be.
That is why in order to provide separability one must put  $\Omega = 0$.
The separation of variables for this case was studied earlier  \cite{Frolov89}.
In the
special case where the string is confined to  the equatorial plane, the
Hamilton-Jacobi equations are trivially separable since the  function
$B(\theta)$
is eliminated from the outset. It is then straightforward to obtain the results
of section \ref{sub4.2}.

\newpage
\centerline{\large{Figure Captions}}

\hspace{0.5cm}

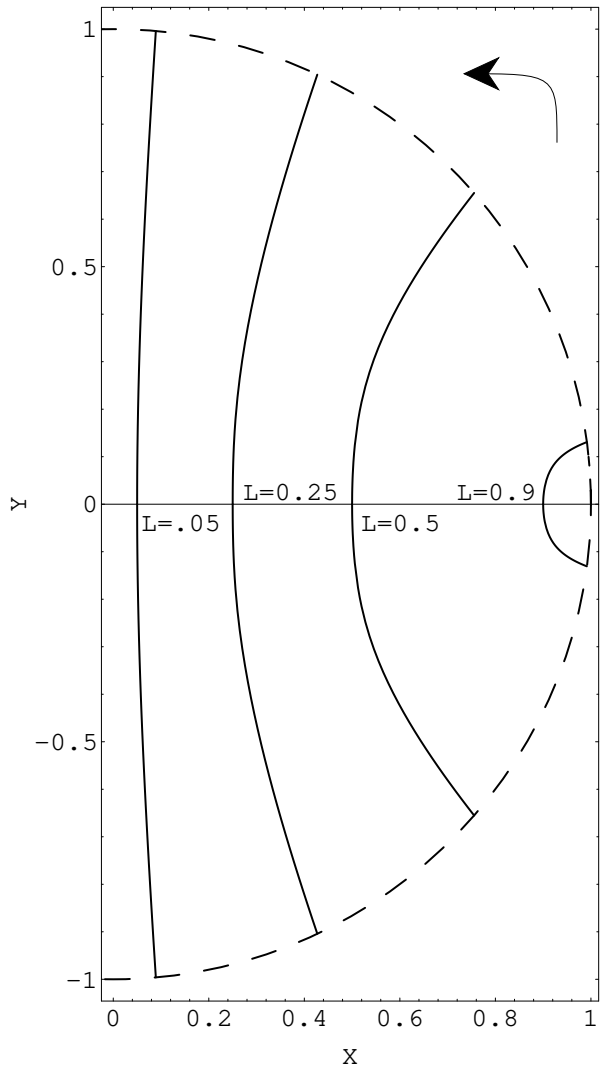
{\bf Figure 1.} String configurations in flat spacetime for $p = 0$ with
$L<1/\Omega$. Solid lines represent strings for 4 different values $L=0.05,
0.25, 0.5,$ and $0.9$ of the angular momentum. A dashed line is a null cylinder
$\rho=1/\Omega$ (here $\Omega=1$). The arrow in this, and subsequent, figures
indicates the direction of rotation of the strings.

\hspace{0.5cm}

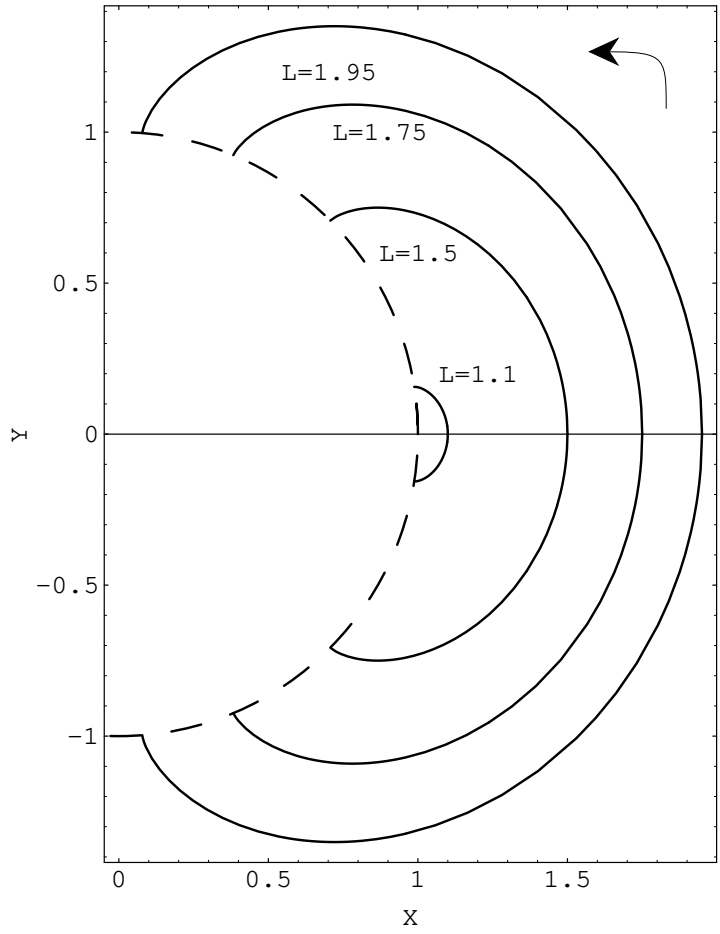
{\bf Figure 2.} String configurations in flat spacetime for $p = 0$ with
$L>1/\Omega$. Solid lines represent strings for 4 different values $L=1.1, 1.5,
1.75,$ and $1.95$ of the angular momentum. A dashed line is a null cylinder
$\rho=1/\Omega$ (here $\Omega=1$).

\hspace{0.5cm}

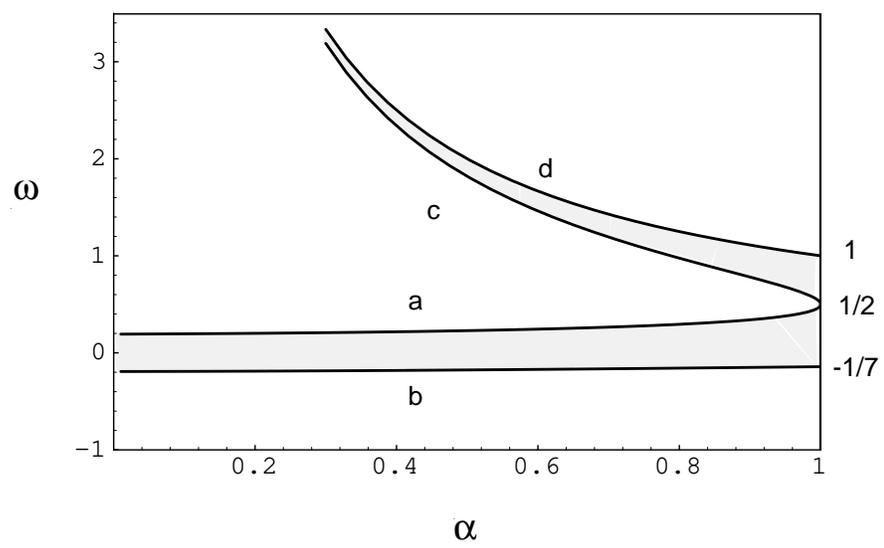
{\bf Figure 3.} The roots of equation (\ref{nn}) are plotted in the
$\alpha-\omega$
plane (curves a, b and c) along with the curve $\omega=1/\alpha$ (curve d).
The shaded regions correspond to parameter values where $f$ has
two positive roots; in the upper region between $c$ and $d$ these roots lie
inside the inner
Cauchy horizon of the black hole and in the lower region between $a$ and $b$
they lie outside
the event horizon of the black hole.

\hspace{0.5cm}

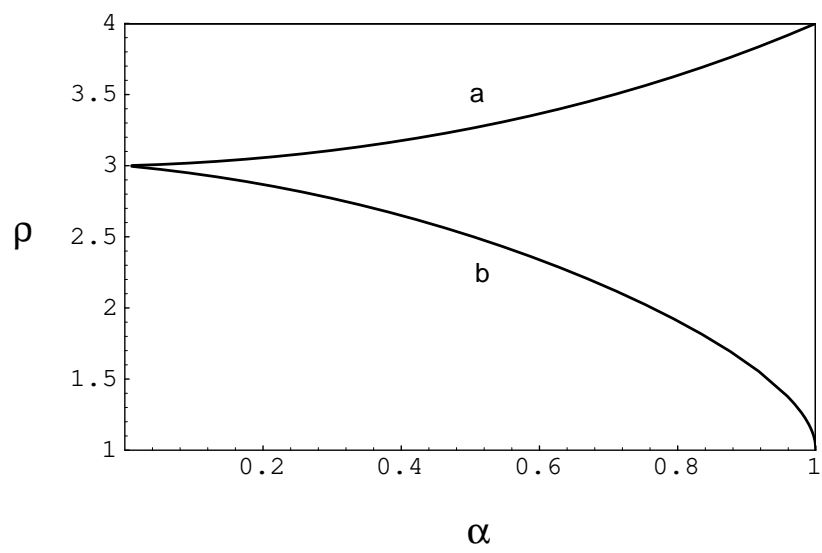
{\bf Figure 4.} The functions $\rho_{max} (\alpha)$ (curve a) and $\rho_{min}
(\alpha)$ (curve b) are plotted for $0<\alpha < 1$.

\hspace{0.5cm}

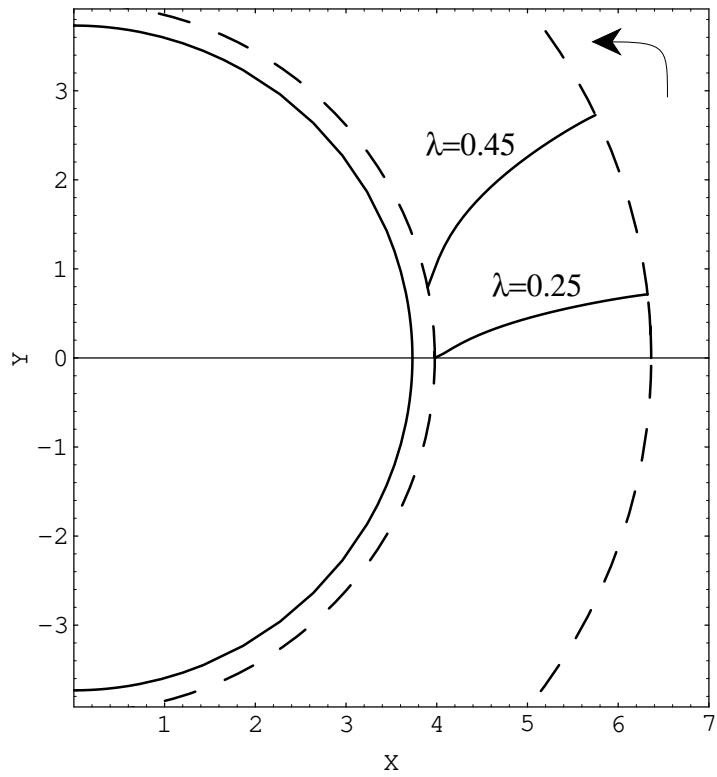
{\bf Figure 5.} This and next figures illustrate qualitatively different types
of motion of rigidly rotating strings in the Kerr spacetime. In all figures the
inner solid circular line is the event horizon $\rho_+$. The nearest to the
horizon dashed circle is $\rho=\rho_1$, and the outer dashed circle (if shown)
is $\rho=\rho_2$. String configurations at the given moment of time are shown
by solid lines with the indication of the corresponding value $\lambda$ of the
angular momentum parameter. The present figure illustrates
Case 1(a) and shows a typical pair of string configurations ($\lambda=0.25$,
$\lambda=0.45$) in the region $\rho_1<\rho < \rho_2$ with no turning points.
The strings have end-points at $\rho=\rho_1$ and $\rho=\rho_2$ ($\alpha=0.5$,
$\omega=0.05$).

\hspace{0.5cm}

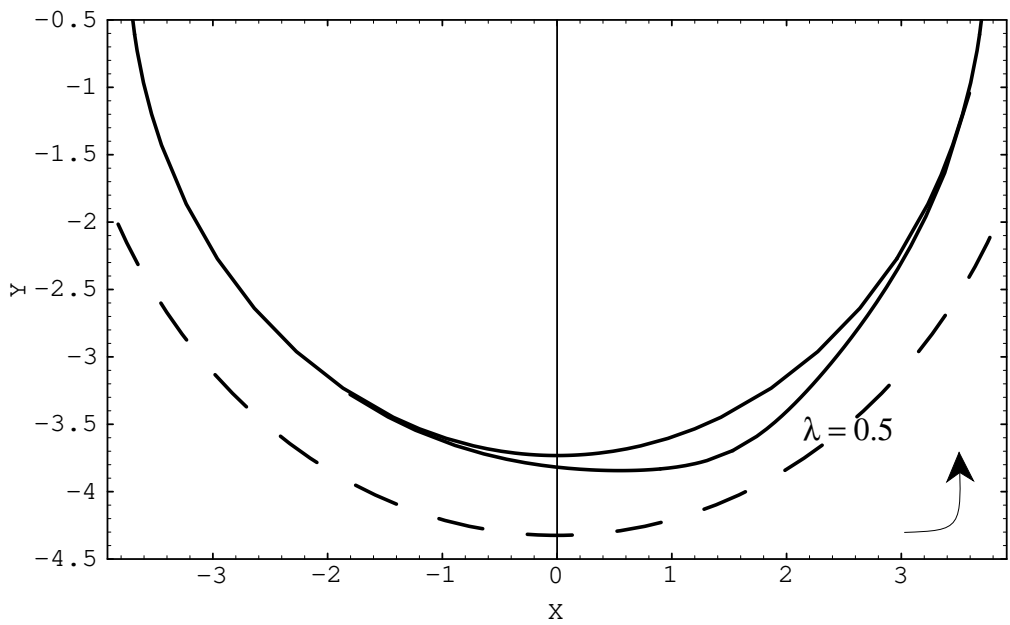
{\bf Figure 6.} Case 1(b): A string configuration ($t=\mbox{const}$ slice) in
the region $\rho < \rho_0$. The string has a turning point at $\rho=\rho_0$
($\alpha=0.5$, $\omega=0.06$).

\hspace{0.5cm}

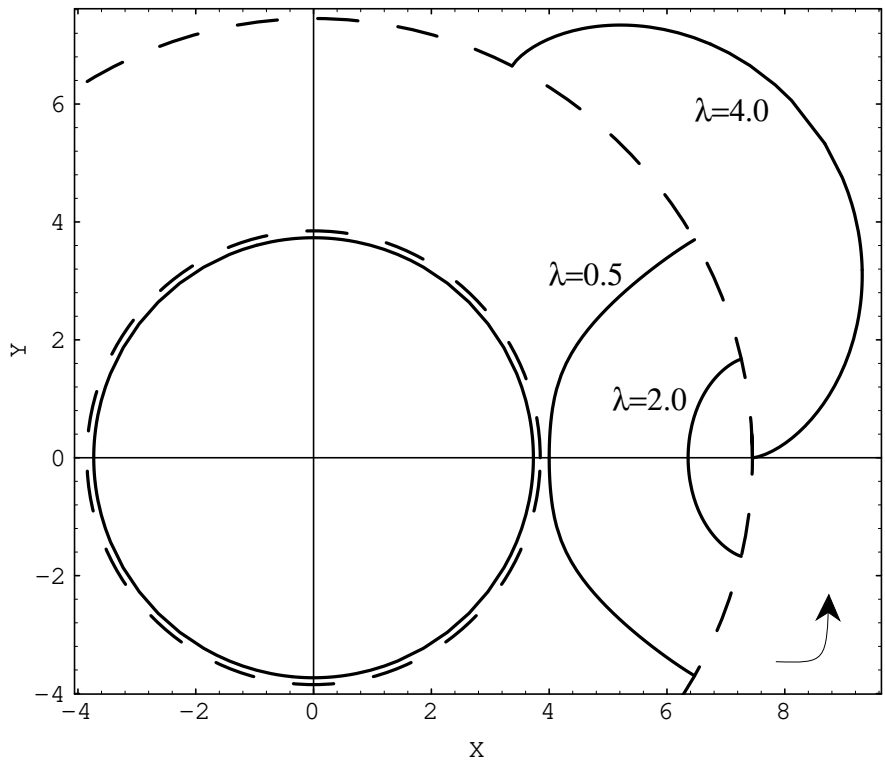
{\bf Figure 7.} Case 2(a): String configurations  inside the region
$\rho_0\le\rho<\rho_2$ ($\lambda=0.5$, $\lambda=2.0$) with turning points at
$\rho_0$. Case 3(b): A  typical string configuration in the region
$\rho_2<\rho\le\rho_0$ ($\lambda=4.0$) with a turning point at $\rho_0$
($\alpha=0.5$, $\omega=0.05$).

\hspace{0.5cm}

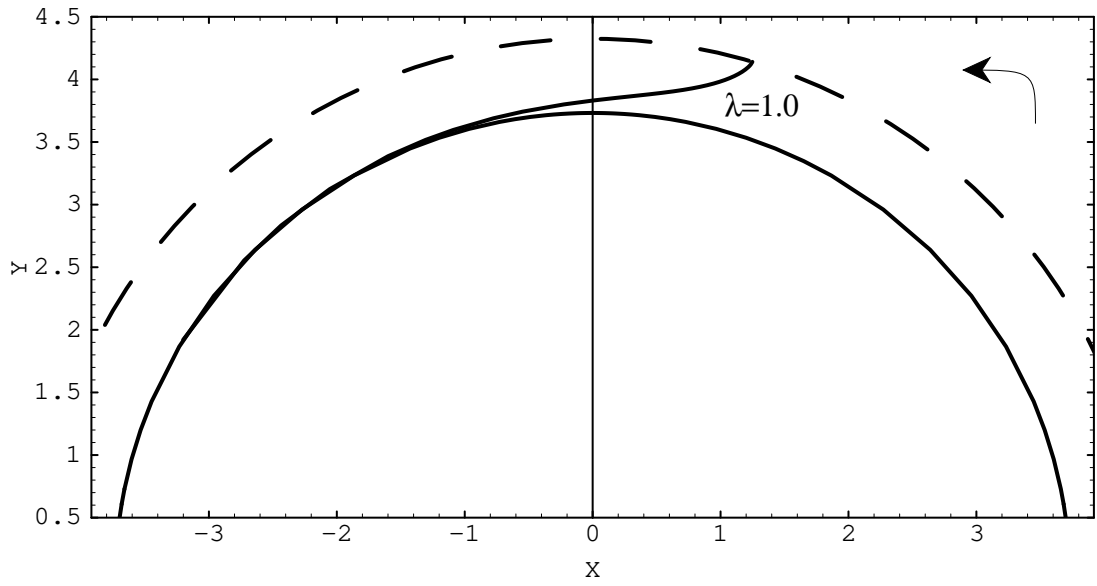
{\bf Figure 8.} Case 2(b): A typical string configuration  where $\rho<
\rho_1$. The string is seen to spiral into the horizon $\rho=\rho_+$ from
$\rho=\rho_1$ ($\alpha=0.43$, $\omega=0.04$).

\hspace{0.5cm}

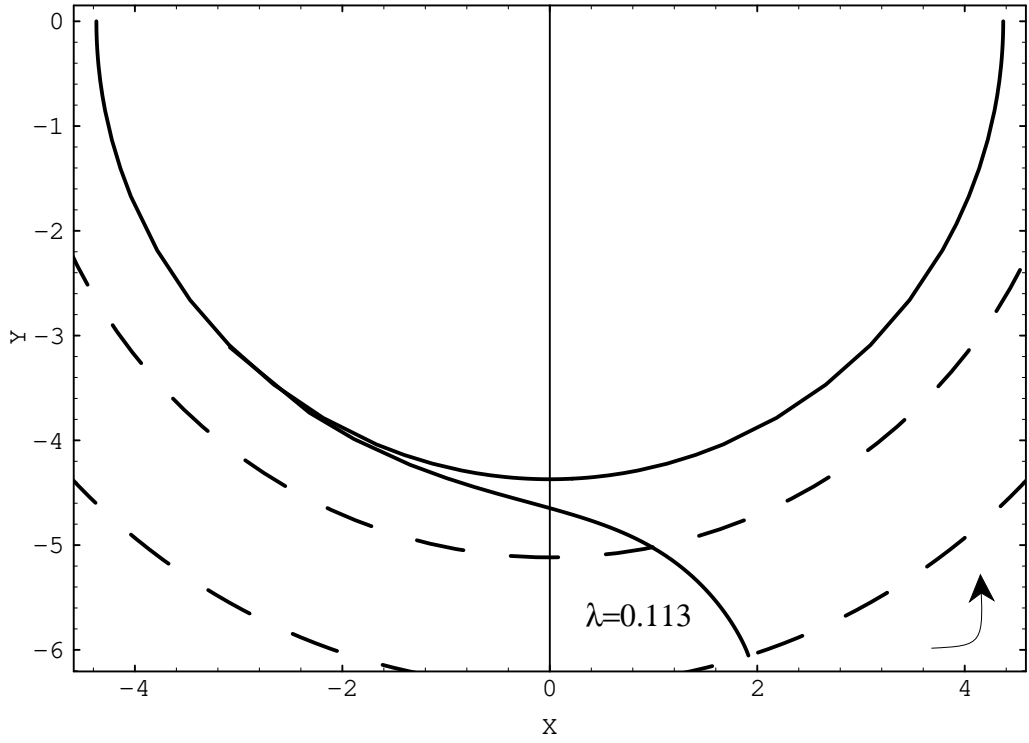
{\bf Figure 9.} A string configuration  where $\rho_0 = \rho_1$. The string is
seen to pass through $\rho_1$ and to spiral into the horizon $\rho=\rho_+$
($\alpha=0.43$, $\omega=0.04$).


\begin{thebibliography}{11}

\bibitem{Vilenkin} A.Vilenkin and P.Shellard, {\it Cosmic Strings and
Other Topological Defects}, Cambridge University Press (1994).

\bibitem{Sanchez} H.J. de Vega and N. Sanchez, Talk given at the 3rd Colloque
Cosmologique, (Paris, France 1995), hep-th/9512074.

\bibitem{Veneziano} N.Sanchez and G.Veneziano, Nucl. Phys. {\bf B333},
253, (1990).

\bibitem{Moss} S.Lonsdale and I.Moss, Nucl. Phys., {\bf B298},
693, (1988).

\bibitem{Us96b} V.P.Frolov, S.C.Hendy and A.L.Larsen, Phys. Rev. {\bf D54},
5093, (1996).

\bibitem{Frolov89} V.Frolov, V. Skarzhinski, A. Zelnikov and O.Heinrich,
Phys. Lett., {\bf B224}, 255, (1989).

\bibitem{Carter89} B.Carter and V.Frolov, Class. Quant. Grav., {\bf 6}:569
(1989).

\bibitem{Carter91} B.Carter, V.Frolov and O.Heinrich, Class. Quant. Grav.,
8:135 (1991).

\bibitem{Wald84} R.M. Wald, {\it General Relativity}, University of Chicago
Press, Chicago, (1984).

\bibitem{deVega96} H.J. de Vega and I.L. Egusquiza, Phys. Rev. {\bf D54}, 7513
(1996).

\end{thebibliography}
\end{document}